\documentclass[aps,groupedaddres8s,fleqn,nofootinbib,twocolumn]{revtex4}
\usepackage{graphicx}
\usepackage{xcolor}
\usepackage{amsmath,amssymb}
\usepackage{float}
\usepackage{physics}
\usepackage{amsfonts}
\usepackage{verbatim}
\usepackage{flushend}
\usepackage{balance}
\usepackage{endnotes}
\usepackage{footnote}
\usepackage{adjustbox}
\usepackage{gensymb}
\usepackage{float}
\usepackage{epigraph}
\usepackage{xcolor}
\usepackage[utf8]{inputenc}
\usepackage{setspace}

\newcommand{\ld}{\lambda_d}
\newcommand{\mfp}{l_{\mathrm{FP}}}

\begin{document}
\title{Double universality of the transition in the supercritical state}
\author{C. Cockrell$^{1}$ and K. Trachenko$^1$}
\address{$^1$ School of Physical and Chemical Sciences, Queen Mary Universiaty of London, Mile End Road, London, E1 4NS, UK}


\begin{abstract}
Universality aids consistent understanding of physical properties. This includes understanding the states of matter where a theory predicts how a property of a phase (solid, liquid, gas) changes with temperature or pressure. Here, we show that the matter above the critical point has a remarkable double universality not limited by pressure and temperature. The first universality is the transition between the liquidlike and gaslike states seen in the crossover of the specific heat on the dynamical length scale in deeply supercritical state and characterised by a fixed inversion point. The second universality is the operation of this effect in many supercritical fluids, including N$_2$, CO$_2$, Pb, H$_2$0 and Ar. Despite the differences in structure and chemical bonding in these fluids, the transition has the same fixed inversion point deep in the supercritical state. This provides new understanding of the supercritical state previously considered to be a featureless area on the phase diagram and a theoretical guide for improved and more efficient deployment of supercritical fluids in green and environmental applications.
\end{abstract}

\maketitle

\section{Introduction}


Our view of the phase diagram of ordinary matter is dominated by the three states of solid, liquid, and gas, and the first order phase transition lines between them which branch out from the triple point. Of these phase transitions, two possess coexistence lines which are finite in length, including the solid-gas sublimation line and the liquid-gas boiling line terminating at the critical point. The matter above the critical point, the supercritical matter, was not thought of as a distinct state of matter, instead seen as a homogeneous state intermediate to liquids and gases and lacking transitions. In particular, distinction between liquidlike and gaslike states within this region was thought to be impossible \cite{landaustat,Kiran2000}. Critical anomalies such as the heat capacity maxima do not persist far beyond the critical point and furthermore depend on the path taken on the phase diagram \cite{Xu2005,Proctor2020a}. Understanding both the supercritical and liquid states involves several fundamental problems related to dynamical disorder and strong inter-molecular interactions \cite{landaustat,Proctor2020a,proctor1,wallacecv,chen-review,Trachenko2016}. Yet such understanding is believed to enhance the deployment of supercritical fluids in important green and environmental applications \cite{Eckert1996,Sarbu2000,Akiya2002,Savage1999,Huelsman2013,Kiran2000}.

The Frenkel line separates two qualitatively dynamical regimes of particle motion: combined oscillatory and diffusive motion below the line and purely diffusive above the line \cite{cockrell2021b}. Practically, the line is calculated from either the dynamical criterion based on the minima of velocity autocorrelation function or the thermodynamic criterion based on the disappearance of transverse modes. This separation of the supercritical state into two different states involves a physical model. It is interesting to ask whether this separation can also be done in a way which is model-free? A related question is whether the separation involves universality across all supercritical systems in terms of suitably identified physical parameters?

Here, we show that deeply supercritical state has a clearly identifiable transition between liquidlike and gaslike states seen in the dependence of the specific heat $c_V$ on the dynamical length $\ld$ which is doubly-universal. The first universality is a fixed path-independent inversion point of the $c_V(\ld)$ crossover, seen as the change of the sign of the derivative of $c_V$ with respect to $\ld$. The second universality is that the location of the inversion point is similar in all simulated fluids, including supercritical N$_2$, CO$_2$, Pb, Ar and to some extent in H$_2$O. Supercritical water has an anomaly, displaying similarities but also differences to the other systems in which the universal transition is identified. The inversion point therefore constitutes a system-independent, path-independent, and an unambiguous separation between two physically distinct supercritical states.

\section{Results and Discussion}

\subsection{Specific heat and dynamical length}

Using molecular dynamics simulations (see the ``Methods'' section for detail), we have simulated several fluids with different structure and chemical bonding in order to ascertain the effect in a wide range of systems. We simulate molecular (N$_2$, CO$_2$), metallic (Pb), hydrogen-bonded network fluid (H$_2$O) and noble Ar. Supercritical CO$_2$ and H$_2$O are particularly important from the industrial point of view due to their deployment in extracting, cleaning, dissolving, environmental and green energy applications \cite{Akiya2002,Savage1999,Huelsman2013,Kiran2000}. We simulate these fluids along several isobars, isotherms, and isochores in the deep supercritical state.

We zero in on the dependence of the specific heat $c_V$ on the dynamical length $\ld=c\tau$, where $\tau$ is liquid relaxation time \cite{Frenkel1955} (for details of calculation and interpretation of $\tau$, see the Methods section) and $c$ is the transverse speed of sound. The specific heat, $c_V$ (heat capacity per atom), is an obvious important choice of a thermodynamic quantity because it reflects the degrees of freedom in the system. The role of the dynamical length $\ld$ is that it sets the upper range of wavelengths of transverse phonons in the liquidlike regime of supercritical dynamics below the Frenkel line (FL) \cite{cockrell2021b}. Details of this mechanism are given in the Methods section. In the gaslike dynamics above the FL, $\ld$ corresponds to the particle mean free path and sets the wavelength of the remaining longitudinal mode. This way, $\ld$ governs the phase space available to phonons in the system. Since the energy of these phonons contributes to liquid $c_V$ \cite{Proctor2020a,proctor1,wallacecv,chen-review,Trachenko2016}, we predict a unique universal relationship between $c_V$ and $\ld$ in the supercritical state.

\begin{figure}
             \includegraphics[width=0.95\linewidth]{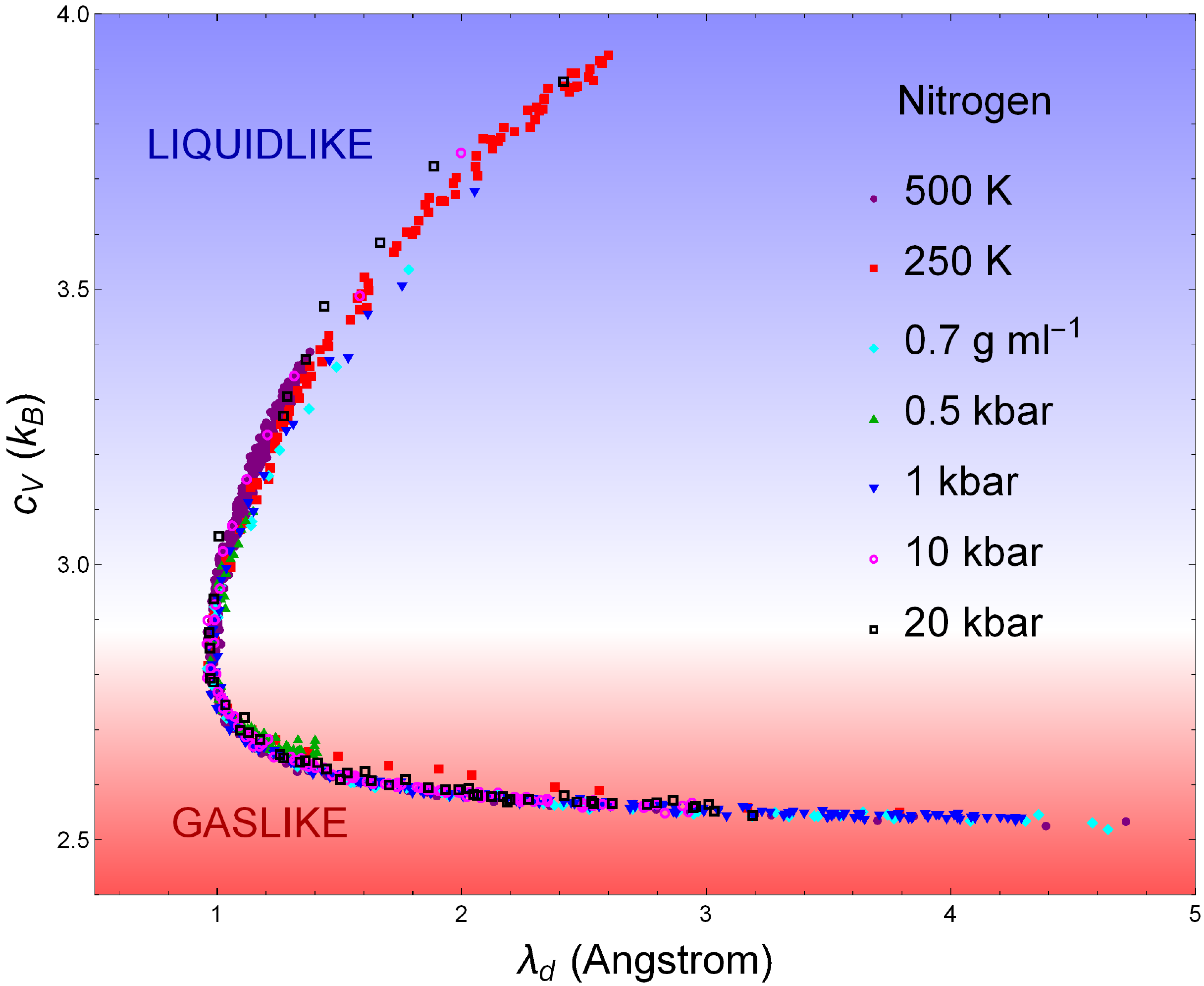}
    \caption{Specific heat $c_V$ in the units of $k_{\rm B}$ as a function of the dynamical length $\ld$ in supercritical nitrogen across 7 phase diagram paths spanning the supercritical state up to 240 times the critical temperature and 3700 times the critical pressure, showing the collapse onto the main sequence with inversion point at  $c_V \approx 2.9 $ and $\lambda_d=1~\rm{\AA}$. Here and elsewhere $k_{\rm B}=1$.}
    \label{fig:n2_ccurves}
\end{figure}

\begin{figure}
             \includegraphics[width=0.95\linewidth]{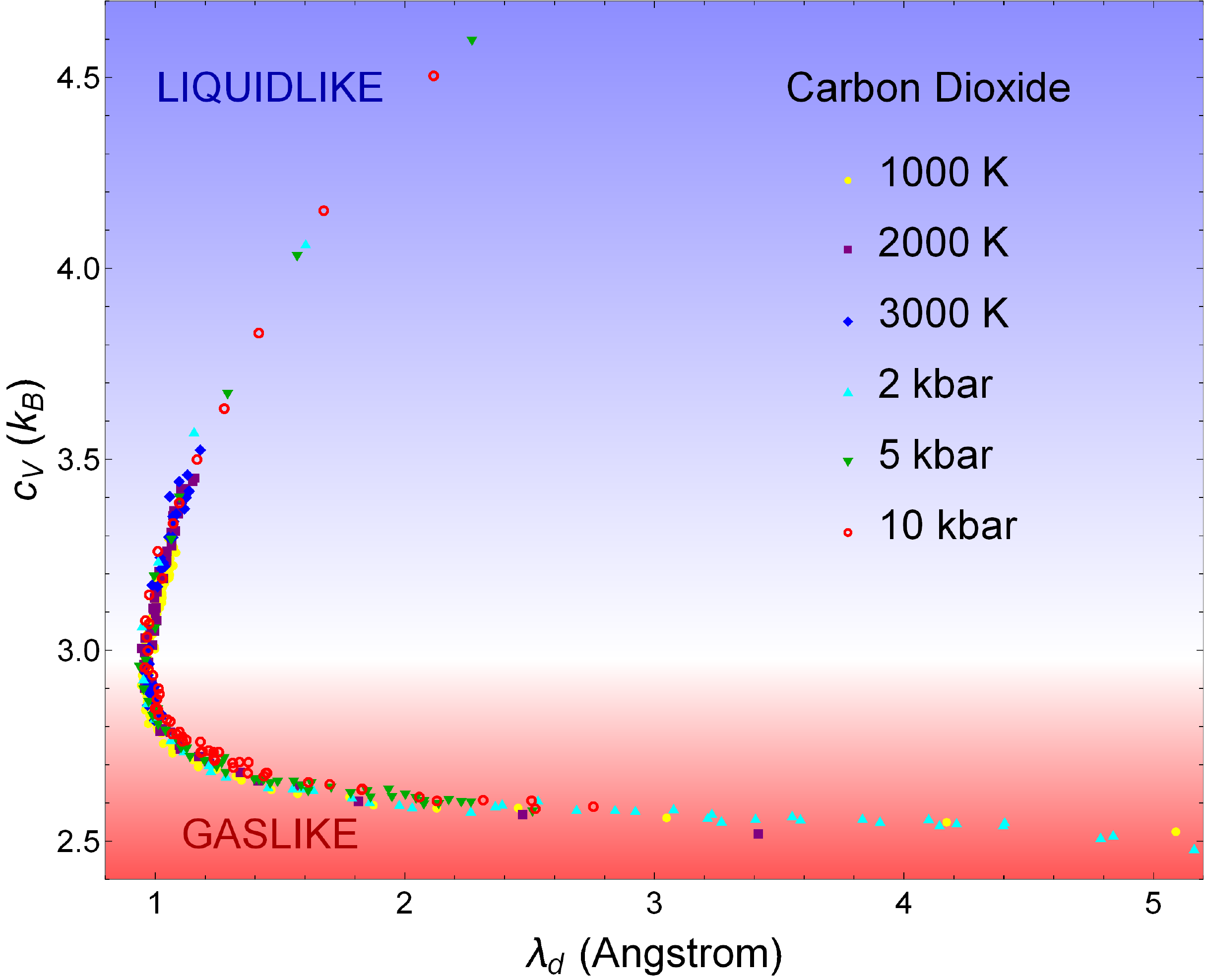}
    \caption{Specific heat $c_V$ as a function of the dynamical length $\ld$ in supercritical carbon dioxide across 6 phase diagram paths spanning the supercritical state up to 33 times the critical temperature and 550 times the critical pressure, showing the collapse onto the main sequence with inversion inversion point at  $c_V \approx 2.9 $ and $\lambda_d=1~\rm{\AA}$.}
    \label{fig:co2_ccurves}
\end{figure}

\begin{figure}
             \includegraphics[width=0.95\linewidth]{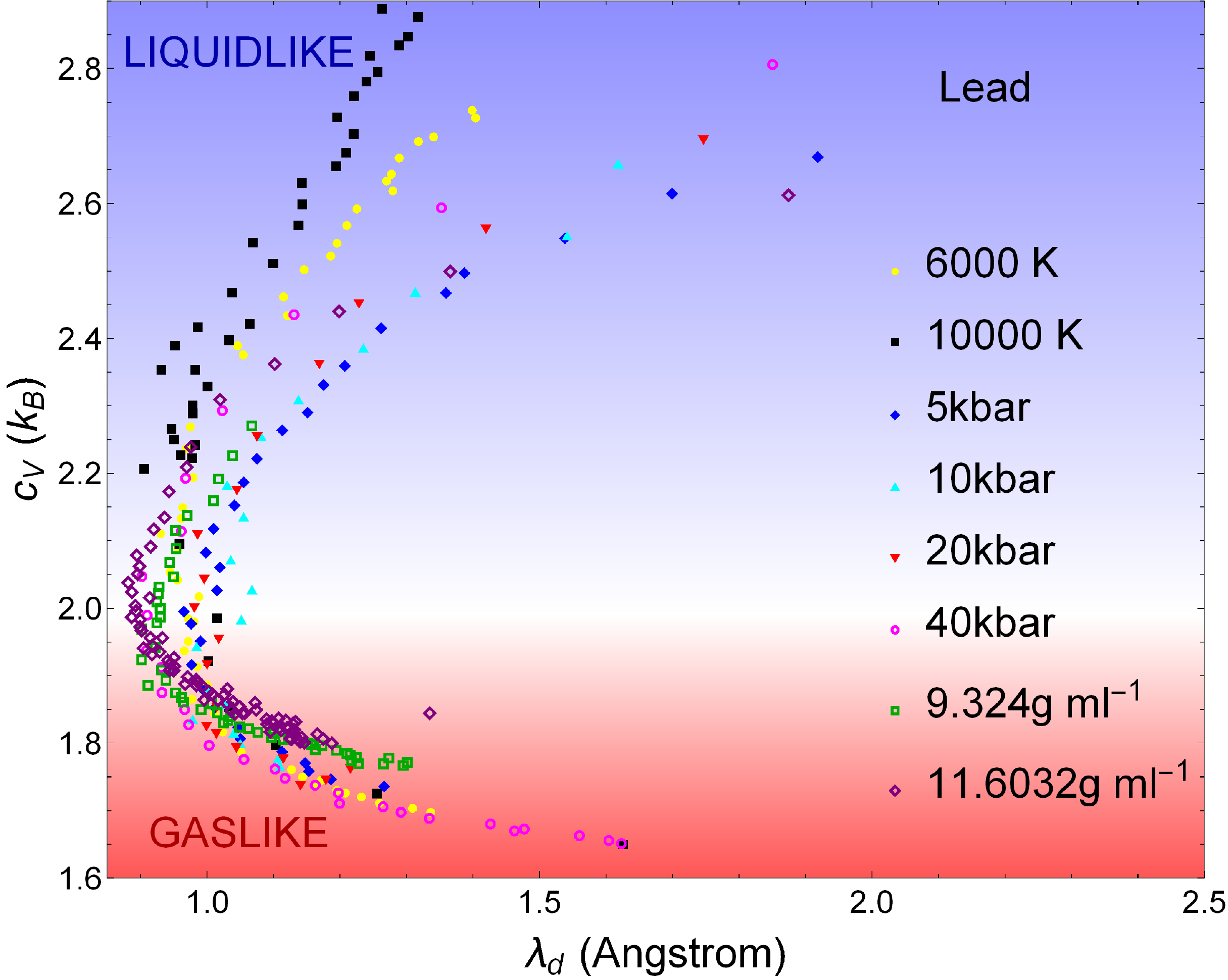}
    \caption{Specific heat $c_V$ as a function of dynamical length $\ld$ in supercritical lead along 8 phase diagram paths spanning the supercritical state up to 20 times the critical pressure and 3400 times the critical temperature, showing the collapse onto the main sequence but with path dependence remaining. The different paths exhibit the same qualitative behaviour and share an inversion point of $c_V \approx 2.0$ and $\ld \approx 1$ \AA.}
    \label{fig:pb_ccurves}
\end{figure}

\begin{figure}
             \includegraphics[width=0.95\linewidth]{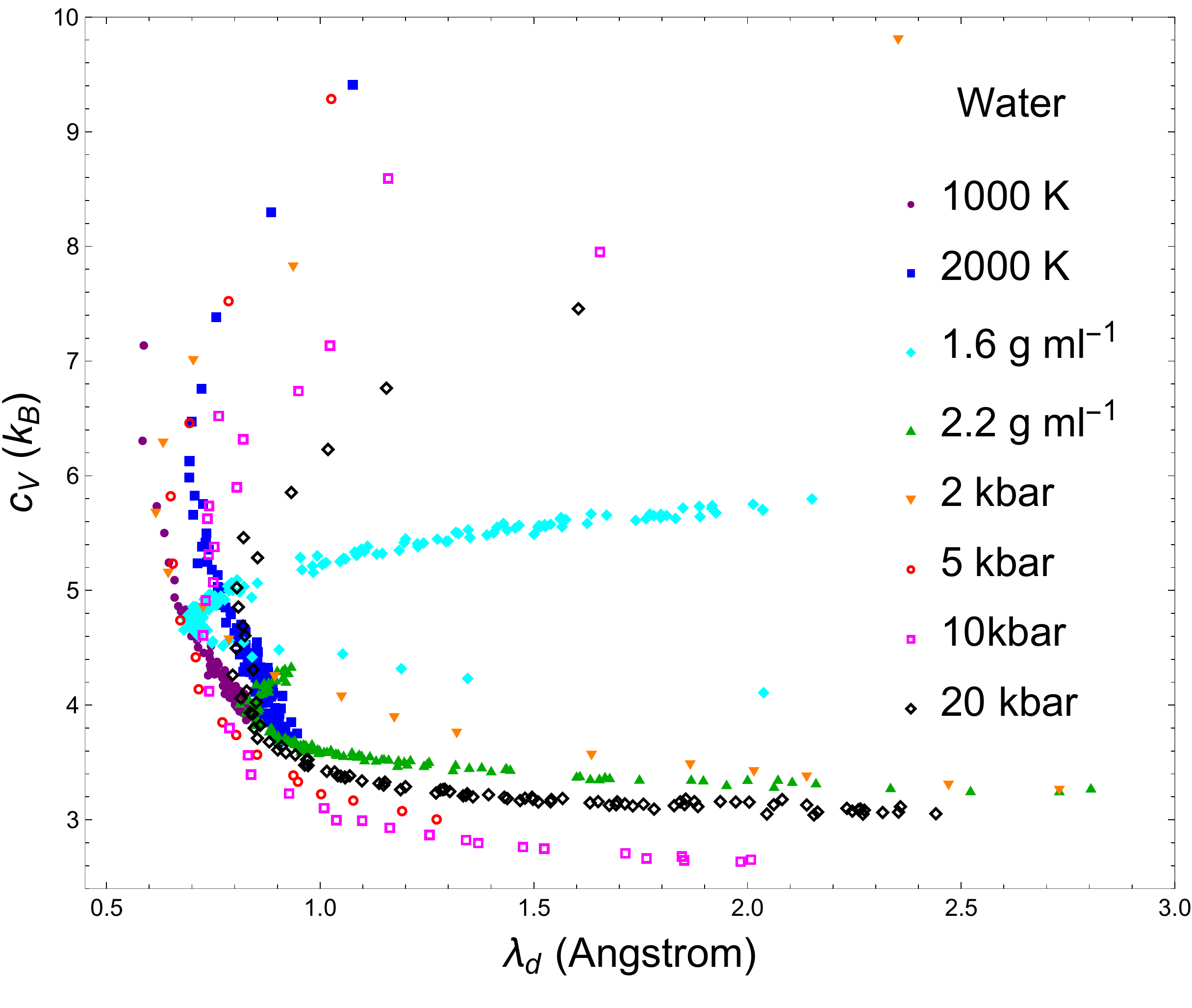}
    \caption{Specific $c_V$ as a function of dynamical length $\ld$ in supercritical water along 8 phase diagram paths spanning the supercritical state up to 15 times the critical temperature and 500 times the critical pressure, showing significant differences in behaviour between different phase diagram paths including the location of the inversion point in $c_V$ and $\ld$.}
    \label{fig:water_ccurves}
\end{figure}

\begin{figure}
             \includegraphics[width=0.95\linewidth]{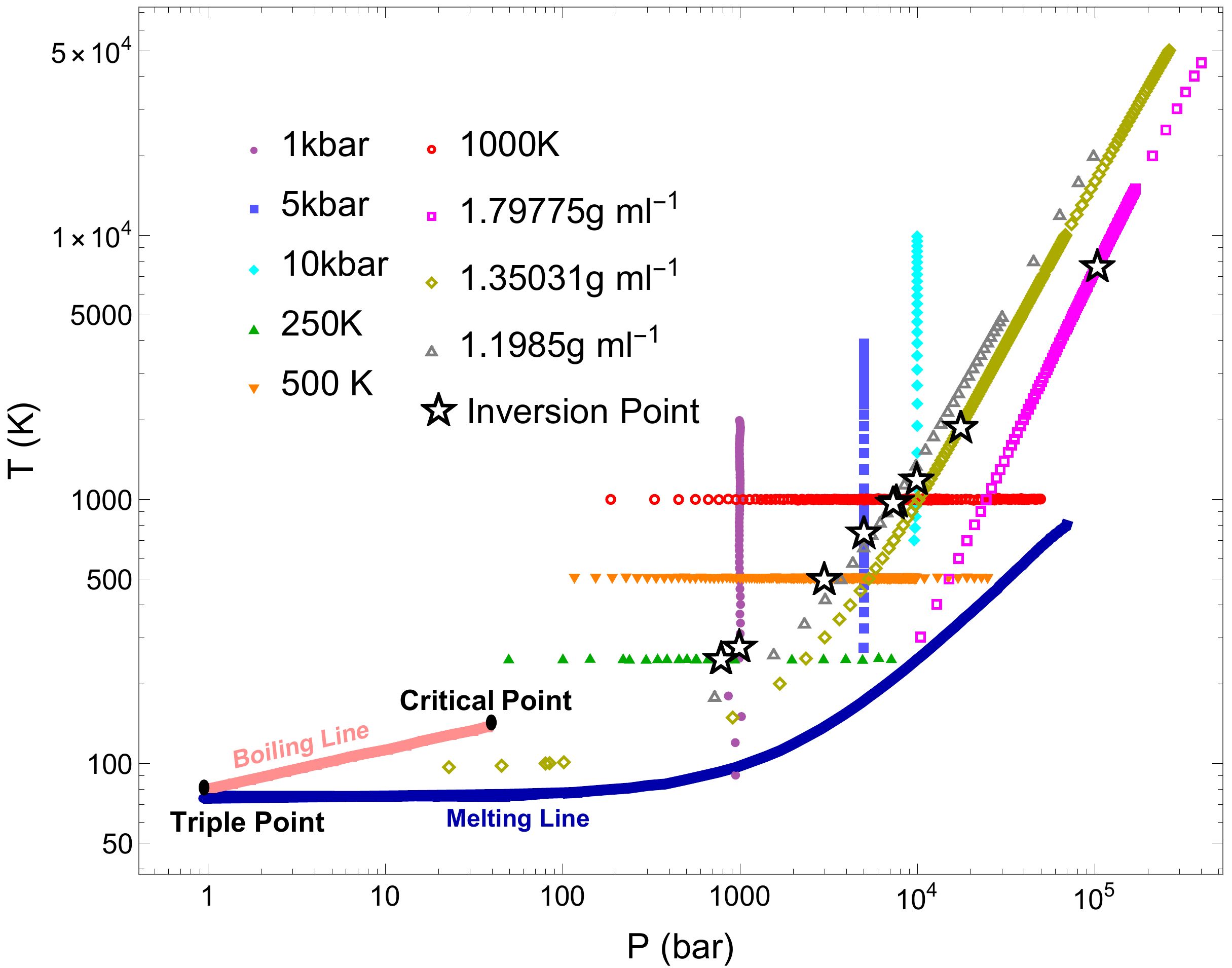}
    \caption{Paths on the phase diagram and simulated pressure and temperature points for argon. Also labelled are the state points where the transition at the inversion point takes place. The triple point, the critical point, together with the boiling and melting lines are shown.
    }
    \label{fig:argon_phasediagram}
\end{figure}

\begin{figure}
             \includegraphics[width=0.95\linewidth]{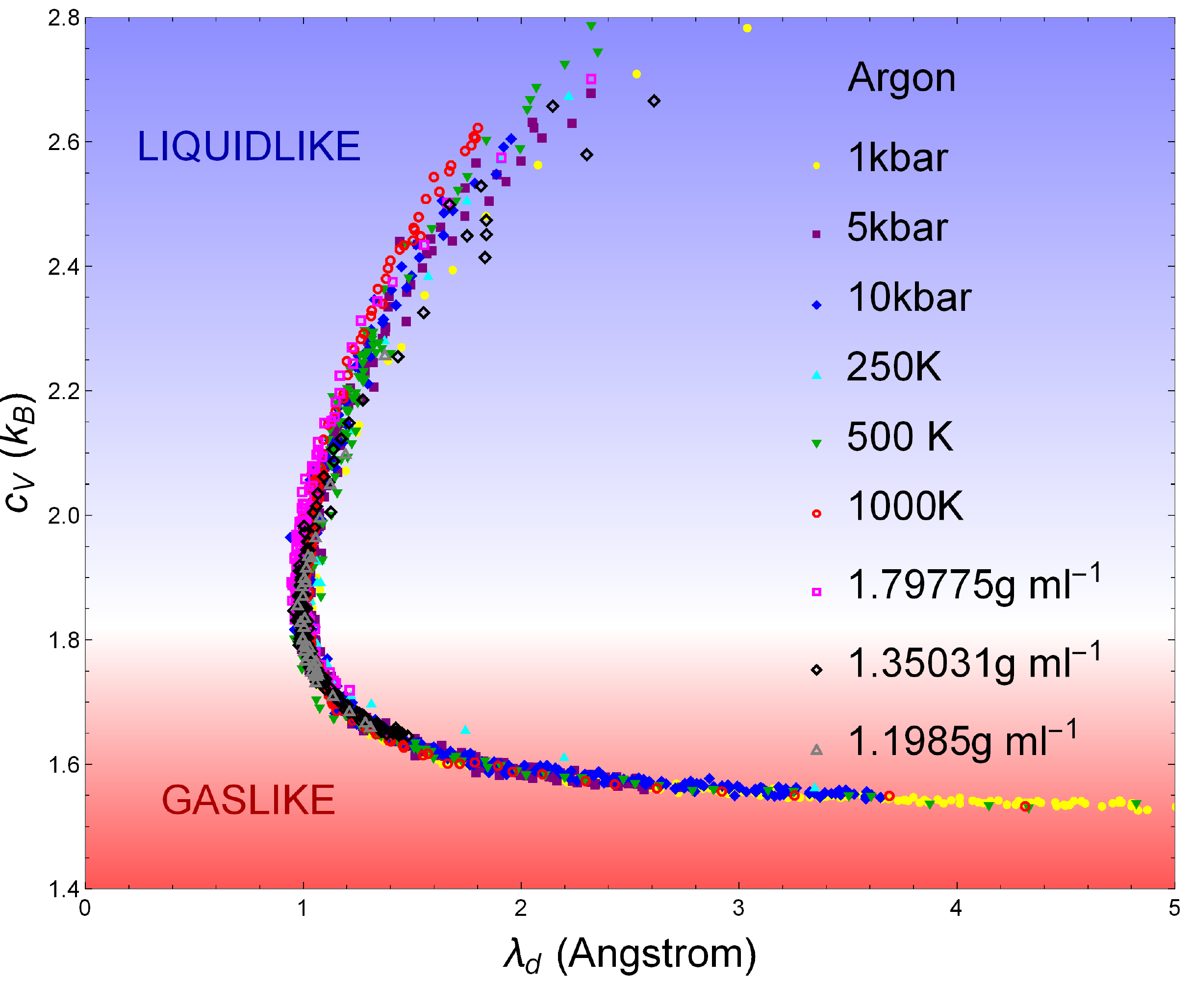}
    \caption{Specific $c_V$ as a function of the dynamical length, $\lambda_d=c\tau$ across 9 paths spanning the supercritical state of argon up to 300 times the critical temperature and 8000 times the critical pressure. The data are from Ref. \cite{Cockrell2021}. All these paths collapse onto the main sequence curve, thereby undergoing a unified dynamic-thermodynamic transition at the path-independent inversion point $c_V \approx 1.9 $ and $\lambda_d=1~\rm{\AA}$.}
    \label{fig:argon_ccurves}
\end{figure}

We show the calculated plots of $c_V$ on $\ld$ in nitrogen, carbon dioxide, lead and water in Figs. \ref{fig:n2_ccurves}-\ref{fig:water_ccurves}. We set $k_{\rm B}=1$ everywhere in the paper. The variation of $c_V$ and $\ld$ shown in these Figures corresponds to a very wide range of pressure and temperature. To illustrate this, we also plot several representative paths simulated on the pressure and temperature phase diagram using the argon data \cite{Cockrell2021} in Fig. \ref{fig:argon_phasediagram} and the corresponding plot $c_V$ vs $\ld$ in Fig. \ref{fig:argon_ccurves}.

The dependence of $c_V$ on $\ld$ across all simulated paths nearly collapses onto a group of ``c"-shaped curves, which we refer to as the \textit{main sequence}. The main sequence is ``c"-shaped and has an inversion point corresponding to the change of the sign of the derivative of $c_v$ with respect to $\ld$.

The origin of the inversion point is as follows. The dynamical length always has a minimum as a function of temperature when crossing from liquidlike to gaslike regimes of particle dynamics. Recall that this crossover is related to the dynamical crossover at the Frenkel line (FL) \cite{cockrell2021b}. In the liquidlike regime below the FL, particle dynamics combines oscillatory motion around quasi-equilibrium positions and flow-enabling diffusive jumps between these positions \cite{Frenkel1955}. In this regime, $\tau$ and $\ld=c\tau$ decrease with temperature. In the gaslike regime above the FL, the oscillatory component of particle motion is lost, leaving the diffusive jumps only \cite{cockrell2021b}. In this regime, $\ld$ becomes the particle mean free path which increases with temperature (see the Methods section for more details). The inversion point is therefore related to the transition between liquidlike and gaslike particle dynamics.

The values of $\lambda_d$ and $c_V$ at the inversion point are physically significant. The value of $\ld = 1~$\AA~corresponds to the ultraviolet cutoff, approximately equal to the shortest length scale in the condensed matter system: the interatomic separation set by the length of the chemical bond. When $\ld$ matches this lengthscale, the fluid stops supporting all transverse phonons simply because the modes with shorter wavelength are non-existent. Concomitantly, particle dynamics can be viewed as the motion with the particle mean free path approximately equal to the interatomic separation.

The value of $c_V$ of about 2 in monatomic argon and lead is important too. $c_V=2$ corresponds to the loss of the contributions from the two transverse phonon branches, with only the kinetic part $\left(c_V=\frac{3}{2}\right)$ and the potential part of the longitudinal mode $\left(c_V=\frac{1}{2}\right)$ remaining. Since this loss corresponds to the disappearance of the oscillatory component of particle motion, $c_V=2$ is taken as a thermodynamic criterion of the FL \cite{Trachenko2016,cockrell2021b}. Phonon anharmonicity can change this result by a relatively small amount \cite{Trachenko2016}, and the disappearance of transverse modes corresponds to $c_V=2$ approximately. The inversion point in nitrogen and carbon dioxide corresponds to $c_V=2.8-2.9$ (in molecular systems, $c_V$ is heat capacity per molecule) due to the additional rotational term contributing $1$ to $c_V$. Subtracting $1$ from calculated $c_V$, we arrive at $c_V=1.8-1.9$ as in monatomic fluids.

We note that the c-plot is not limited in pressure and temperature as long as the system remains chemically unaltered (the same proviso as for the melting line), extending to the entire supercritical state of matter.

All phase diagram paths of different types in argon, nitrogen and carbon dioxide collapse onto the main sequence curve. As discussed in the next section in more detail, $c_V(\ld)$ along different phase diagram paths follows the same ``c"-shape of the main sequence in lead but moderate path dependence remains far from the inversion point. This could be related to the electronic contribution (not accounted for in the theory based on phonons) represented by the many-body empirical potential in classical MD simulations.

Water, however, shows a different behavior in Figure \ref{fig:water_ccurves}. This is not unexpected, given that water possesses many anomalies which continue to inspire enquiry and research \cite{Poole1992,Gallo2016}. Water's supercritical state is little understood despite extensive exploitation in industrial and environmental applications \cite{Akiya2002,Savage1999,Huelsman2013,Kiran2000}. The specific heat of liquid water at the melting point at atmospheric pressure is almost twice as high as that of ice and is related to large ``configurational" contribution to the liquid heat capacity. This contribution is related to water-specific hydrogen-bonded network undergoing the coordination change from 4 to 6, with the associated contribution to entropy and specific heat \cite{Eisenberg2005}. This effect precludes the description of water's heat capacity using phonons only as discussed earlier. Although different paths still result in the c-shaped curves, we see significant path dependence in Fig. \ref{fig:water_ccurves}. $c_V$ at the inversion point varies in the range of about 5-6 per molecule. This higher $c_V$ can be understood as a result of the additional configurational term in water mentioned earlier as well as the rotational term. Nevertheless, the inversion points corresponds to $\lambda_d$ close to 1~\AA~as in previous fluids.

\subsection{Path dependence}

\begin{figure}[H]
             \includegraphics[width=0.9\linewidth]{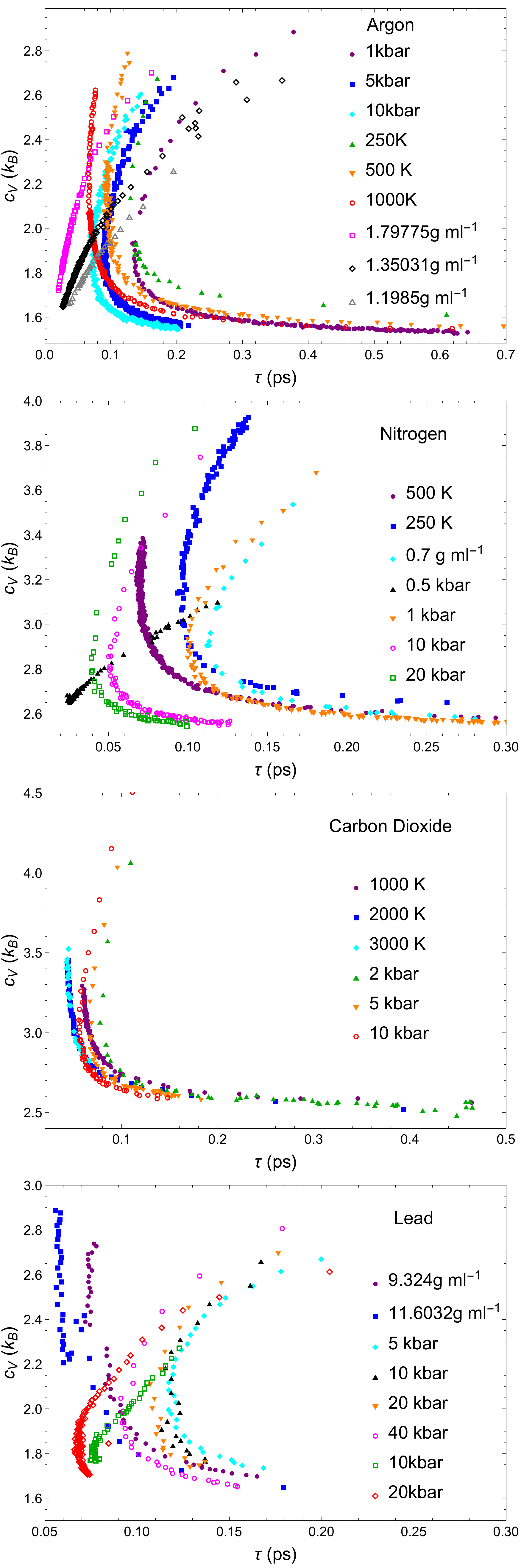}
    \caption{$c_V$ of simulated Ar, N$_2$, CO$_2$ and Pb as a function of relaxation time $\tau$ across different diagram paths.}
    \label{cvtau}
\end{figure}

The universality of the inversion point and ``c"-shaped main sequence curves observed in the previous section is best taken in the context of the path \textit{dependence} of $c_V$ as a function of parameters other than the dynamical length $\lambda_d$. In this study we performed simulations along isobars, isotherms, and isochores. The dynamical parameter $\tau$, the relaxation time introduced in the main article, provides a way to compare $c_V$ vs $\tau$ along different paths. In Fig. \ref{cvtau} we observe substantial path dependence of $c_V(\tau)$ which manifests in several different ways. The first is that different paths, particularly isochores, have different shapes from one another. The second is that these curves do not coincide at the values of $c_V$ or $\tau$. Third, there is no fixed inversion point. This is to be contrasted to the main sequence curves seen in Figures \ref{fig:n2_ccurves}-\ref{fig:water_ccurves} and \ref{fig:argon_ccurves}, wherein all phase diagram paths share a cross-system universal fixed inversion point at the minimal value of about $\lambda_d$ = 1 \AA\ and $c_V=2$ ($c_V=2$ in monatomic fluids or appropriately modified $c_V$ in molecular fluids).

In summary, we see very different plots depending on which path on the phase diagram is chosen: there is no fixed inversion point, and all curves are far away from each other. This variation is removed once we plot $c_V$ vs $\ld=c\tau$ as seen in Figures \ref{fig:n2_ccurves}-\ref{fig:water_ccurves} and \ref{fig:argon_ccurves}.

Similarly to $\ld$, $\tau$ is a dynamical parameter. However the stark path dependence of $c_V(\tau)$ emphasises that the c-transition is not a consequence of simply reducing $c_V$ to dynamics, but that the introduction of the special new dynamical parameter $\lambda_d=c\tau$ is necessary to achieve a fixed inversion point, data collapse and observe double universality discussed in the next section.

We observe that although there is a moderate path dependence of $c_V(\ld)$ for lead in Figure \ref{fig:pb_ccurves}, this path dependence of $c_V(\tau)$ in Figure \ref{cvtau} is much more profound. Hence in cases where the $c_V(\ld)$ plot does not achieve the full data collapse, it brings the paths significantly closer together.

\subsection{Double universality}

We now come to the main finding of this work related to double universality of the c-transition. The {\it first} universality is that for each system, the c-transition plot has an inversion point which is fixed and corresponds to about $\lambda_d=1$~\AA~and $c_V=2$ ($c_V=2$ for monatomic systems or appropriately modified $c_V$ in molecular systems) for all paths on the phase diagram, including isobars, isochores and isotherms, spanning orders of magnitude of temperature and pressure. This inversion point provides an unambiguous, theory-independent and path-independent, transition between liquidlike and gaslike states in the sense discussed earlier. The {\it second} universality is that this behavior is generic on the supercritical phase diagram and is the same for all fluids simulated.

\begin{figure}
                \includegraphics[width=0.98\linewidth]{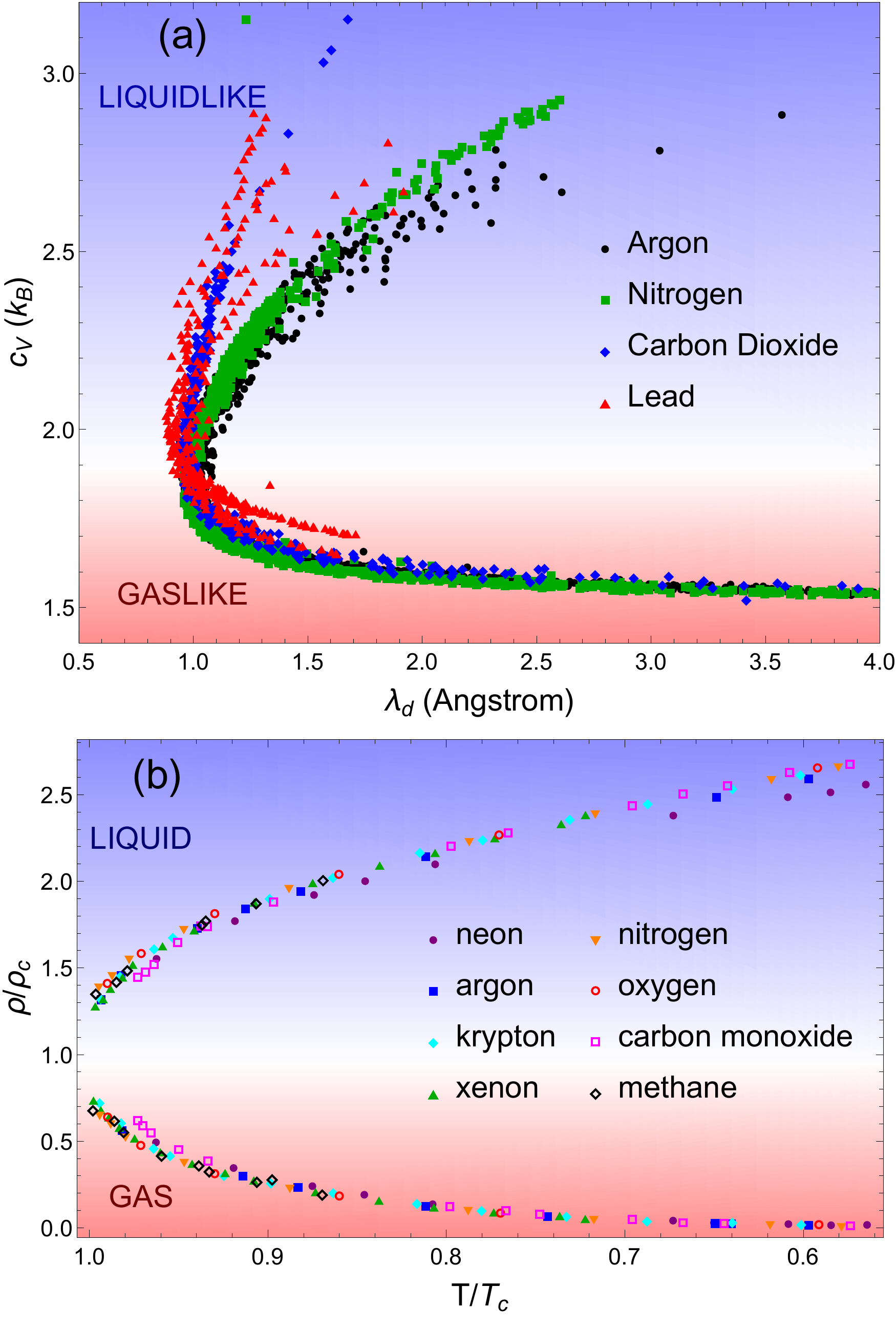}
    \caption{(a) $c_V$ as a function of the dynamical length, $\ld$ across different phase diagram paths and across four different fluids (Ar, CO$_2$, N$_2$ and Pb), showing the system-independent universal inversion point of $c_V \approx 2$, $\ld = 1$~\AA ~(molecular CO$_2$ and N$_2$ have rotational degrees of freedom removed from its $c_V$); (b) (reproduced from Ref. \cite{Guggenheim2004}) the reduced density ($\rho_c$ is the critical concentration) of coexisting liquids and gases at the boiling line at temperatures close to the critical temperature, $T_c$, in a variety of different systems, with all curves coinciding into the same shape.}
    \label{fig:universal_ccurves}
\end{figure}

We now analyse our four systems (which excludes water) on the same set of axes. In order to compare, we must remove the rotational degrees of freedom from the heat capacity of nitrogen and carbon dioxide, which amounts to subtracting 1 from $c_V$ as mentioned earlier. This inter-system plot is presented in Fig. \ref{fig:universal_ccurves}a.

The four fluids exhibit qualitatively similar main sequence curves: the ``c"-shape is present in all curves, and the divergent liquidlike branches converge into almost the same gaslike branch. This plot exhibits what we are calling ``double universality": the function $c_V(\ld)$ across not only different phase diagram paths but also across \textit{different fluids} converges at the universal inversion point of $c_V \approx 2$, $\ld \approx 1$~\AA. This inversion point therefore constitutes a system independent, path independent, and unambiguous model-free separation between liquidlike and gaslike states in the supercritical state.


To draw the analogy with ordinary phase transitions, we recall the behaviour of liquid and gas densities on the coexistence line as the critical point is approached, depicted in Fig. \ref{fig:universal_ccurves}b. The experimental relationship between reduced density and reduced temperature of the coexisting liquids and gases is system-independent for several small noble and molecular elements near the critical point \cite{Guggenheim2004}. In this plot, the specific microscopic details of different systems are often irrelevant to the qualitative behaviour near a phase transition, and the transition falls into a universality class determined by system symmetries and dimensionality \cite{sethna}.

The plots in Fig. \ref{fig:universal_ccurves}a depict the relationship between a thermodynamic quantity, $c_V$, and a dynamical quantity, $\ld$. It is in this sense that we consider the c-transition to represent a dynamical-thermodynamic transition. The system independence of the main sequence for simple fluids is further suggestive of a universal transition operating in the supercritical state. The fixed point of this dynamical-thermodynamic transition approximately corresponds to ($\ld = 1$ \AA, $c_V=2$).

We also note that the ``c"-transition is not observed in proximity to the critical point. The critical anomalies, caused by diverging correlation lengths, present in this region disrupt the relationship between the dynamical length and the heat capacity in all systems studied here. As mentioned earlier, the inversion point is far above the critical.

Finally, the universal inversion point and the related dynamical transition at the FL corresponds to the solubility maxima (known as ``ridges'') and optimal extracting and dissolving abilities of supercritical fluids \cite{cockrell2021b}. This importantly addresses the widely-held belief that improved and more efficient deployment of supercritical fluids will benefit from better theoretical understanding of the supercritical state \cite{Eckert1996,Sarbu2000,Kiran2000}. Our current results therefore give a universal way to locate the inversion point where the performance of a supercritical fluid is optimised, improving the supercritical technologies.

\section{Conclusions}

We have shown that the supercritical state has a remarkable double universality. First, the transition between the liquidlike and gaslike states is characterised by fixed inversion point and near path-independence. Second, this effect universally applies to many supercritical fluids. This provides new understanding of the supercritical state of matter and a theoretical guide for improved deployment of supercritical fluids in green and environmental applications.

\section{Methods}

\subsection{Simulation details}

We use DL\_POLY molecular dynamics simulations package \cite{dlpoly}. For argon and nitrogen, we use the Lennard-Jones potential fitted to their properties. For nitrogen, we use a rigid two-site Lennard-Jones potential \cite{Powles1976}. The potential for carbon dioxide is a rigid-body non-polarisable potential based on a quantum chemistry calculation, with the partial charges derived using the distributed multipole analysis method \cite{Gao2017}. The potential was derived and tuned using a large suite of energies from {\it ab initio} density functional theory calculations of different molecular clusters and validated against various sets of experimental data including phonon dispersion curves and $PVT$ data. These data included solid, liquid and gas states, gas-liquid coexistence lines and extended to high-pressure and high-temperature conditions \cite{Gao2017}. The potential used for water was TIP4P/2005 potential, which is optimised for high pressure and temperature conditions \cite{Abascal2005}. A careful analysis \cite{Vega2009,Vega2011} assigned this potential the highest score in terms of the extent to which the results agree with different experimental properties, including the equation of state, high pressure and temperature behaviour, and structure. The electrostatic interactions were evaluated using the smooth particle mesh Ewald method in the MD simulations of carbon dioxide and water. The potentials for water and carbon dioxide are rigid body potentials. Simulations of lead were performed using an embedded atom model (EAM) potential \cite{Belashchenko2017}, which has been used to calculate the properties of molten lead at temperatures up to 25000 K and 280 GPa, which include the range discussed here. 

Systems are simulated along several isobars, isotherms, and isochores in the deep supercritical state, with all paths but named exceptions being far from the critical point and Widom line \cite{Xu2005} of their respective phase diagrams. Equilibration was performed in the NPT ensemble with the Langevin thermostat in order to generate the mean densities along the isobars and isotherms. For argon, system sizes between 500 and 108000 atoms were used with no discrepancy in calculated quantities, consistent with the earlier ascertained insensitivity of viscosity to system size \cite{Yeh2004}. System sizes of 512 molecules were used for water, nitrogen and carbon dioxide simulations, and 5120 atoms for lead. The timestep used was 1 fs for water and carbon dioxide and 0.5 fs for lead, which conserved total energy under the Velocity-Verlet integrator in the NVE ensemble to one part in $10^5$. Configurations at the target densities on all paths were then generated, which were then equilibrated with the NVT ensemble for 50 ps. Following this equilibration, we generated 20 independent initial conditions for each state point using seeded velocities, and each of these initial conditions were run for 1 ns in the NVE ensemble during which all properties were calculated. We calculated $c_V$ in the NVE ensemble as \cite{Allen1991}:

\begin{equation}
    \label{eqn:cvnve}
    \langle K^2 \rangle - \langle K \rangle^2 = \frac{f}{2} N T^2 \left( 1- \frac{f}{2 c_V}\right)
\end{equation}
with $K$ the kinetic energy, and $f$ the number of translational and rotational degrees of freedom available to the molecule in question.

The shear modulus at high frequency and shear viscosity were calculated using the molecular stress autocorrelation function, from Green-Kubo theory \cite{Zwanzig1965, Balucani1994}:
\begin{equation}
    \label{eqn:GKshearmod}
    G_\infty = \frac{V}{T} \langle \sigma^{x y}(0)^2 \rangle
\end{equation}

\begin{equation}
    \label{eqn:GKviscosity}
   \eta = \frac{V}{T} \int_0^{\infty} \mathrm{d} t \ \langle \sigma^{x y}(t) \sigma^{x y}(0) \rangle
\end{equation}

\noindent with $\sigma^{x y}$ an off-diagonal component of the microscopic stress tensor. The integration of the long-time tails of autocorrelation functions was implemented using the Green-Kubo formulae \cite{Zhang2015}. The 20 independent initial conditions were used to average the autocorrelation function $\langle \sigma^{x y}(t) \sigma^{x y}(0) \rangle$ over these initial conditions. The end result for viscosity was insensitive to adding more initial conditions.

The dynamical length $\lambda_d$ was calculated as $\lambda_d=c\tau$, where $\tau=\frac{\eta}{G_\infty}$, $c^2=\frac{G_\infty}{\rho}$ and $\rho$ is density.

\subsection{Theory: specific heat and dynamical length}

In this section, we explain the physical origin of the inter-relationship between the specific heat and the dynamical length in the supercritical state. The specific heat, $c_V$, is an obvious important choice of a thermodynamic quantity because it reflects the degrees of freedom in the system. The dynamical length and its role are discussed below.

The choice of the dynamical parameter is informed by the Maxwell-Frenkel viscoelastic theory \cite{Maxwell1867, Frenkel1955}. A liquid has a combined response to shear stress:
\begin{equation}
    \label{eqn:viscoelastic}
    \dv{s}{t} = \frac{\sigma}{\eta} + \frac{1}{G_{\infty}} \dv{\sigma}{t}
\end{equation}
where $s$ is the shear strain, $\sigma$ is the shear stress, $\eta$ is the shear viscosity and $G_{\infty}$ is the high-frequency shear modulus. When the external perturbation stops, the internal stress relaxes according to:
\begin{equation}
    \label{eqn:stressrelaxation}
    \sigma(t) = \sigma_0 \exp\left(-\frac{t}{\tau}\right)
\end{equation}
having introduced the Maxwell relaxation time $\tau$:
\begin{equation}
    \label{eqn:taum}
    \tau = \frac{\eta}{G_{\infty}}
\end{equation}

Frenkel related this time to the average time between molecular rearrangements. This relationship is backed up by experiments and modelling \cite{Jakobsen2012,Iwashita2013} and has become an accepted view \cite{Dyre2006}.

Using Eq. \eqref{eqn:viscoelastic}, the Navier-Stokes equation can be generalised to include the elastic response of the liquid, yielding \cite{Trachenko2016}:

\begin{equation}
    \label{eqn:mnavierstokes}
    c^2 \pdv[2]{v}{x} = \pdv[2]{v}{t} + \frac{1}{\tau}\pdv{v}{t}
\end{equation}

\noindent where $v$ is the transverse velocity field, $c$ is the transverse speed of sound $c = \sqrt{G_{\infty}/\rho}$ and $\rho$ the density.

Seeking the solution of Eq. \eqref{eqn:mnavierstokes} in the form $v = v_0 \exp\left( i \left(\omega t - k x\right) \right)$ gives
\begin{equation}
    \label{eqn:omegasol}
    \omega = -\frac{i}{2 \tau} \pm \sqrt{c^2 k^2 - \frac{1}{4 \tau^2}}
\end{equation}

For $k \leq \frac{1}{2 c \tau}$, $\omega$ has no real solutions. For larger $k$, the plane waves decay according to the decay time $\tau$. We therefore define $k_g$ as

\begin{equation}
    \label{eqn:kg}
    k_g = \frac{1}{2 c \tau}
\end{equation}

\noindent which sets the shortest wavevector for propagating transverse phonons and corresponds to the gap in the phonon momentum space \cite{Baggioli2020}.

Here, we work in terms of the ``dynamical length" featuring in Eq. \eqref{eqn:kg}, $\ld$:

\begin{equation}
    \label{eqn:ld}
    \ld = c \tau
\end{equation}

$\ld$ sets the propagation range, or mean free path of transverse phonons, in the liquidlike regime below the FL \cite{cockrell2021b}. This is seen from Eq. \eqref{eqn:omegasol} which gives the decay factor $\exp\left(-\frac{t}{2\tau}\right)$. Since $\tau$ sets the time over which the shear stress decays in the liquid as discussed earlier or, in other words, the lifetime of transverse phonons, $c\tau$ is a measure of their mean free path. This implies no phonons with wavelengths longer than the propagation range and is consistent with Eq. \eqref{eqn:kg}.

In the gaslike regime of particle dynamics above the FL \cite{cockrell2021b}, $\ld$ corresponds to the mean free path of particle motion, $\mfp$ \cite{Frenkel1955}. Indeed, the shear modulus in a fluid with no interactions is $G_\infty=n T$ \cite{Zwanzig1965}, where $n$ is the concentration. Meanwhile, the gaslike viscosity is \cite{Blundell2010} $\eta=\frac{1}{3}\rho v_{\mathrm{th}}\mfp$, where $v_{\mathrm{th}}$ is the thermal velocity and $\mfp$ is the particle mean free path. Noting that $\frac{1}{2} \rho v_{\mathrm{th}}^2=\frac{3}{2}nT$, we find $\tau=\frac{\eta}{G_\infty}=\frac{\mfp}{v_{\mathrm{th}}}$. The dynamical length in the gaslike state is $\ld=c\tau=v_\mathrm{th}\tau$ (since the speed of sound in the liquidlike state below the FL, $c$, approximately becomes thermal velocity of particles in the gaslike state above the FL, $v_\mathrm{th}$), and $\ld=\mfp$.

We can now see the role played by the dynamical length in the liquidlike and gaslike regimes of the supercritical state. $k_g$ in Eq. \eqref{eqn:kg} increases with temperature because $\tau$ decreases, and the dynamical length in \eqref{eqn:ld} becomes shorter. This reduces the phase space available for transverse phonons \cite{Trachenko2016}. When $\tau$ approaches its shortest value comparable to the Debye vibration period $\tau_{\rm D}$, $k_g$ approaches the Brillouin zone boundary because $c\tau_{\rm D}=a$, where $a$ is the interatomic separation. At this point, all transverse modes disappear, corresponding to $c_V=2$ (this value is equal to the kinetic term $\frac{3}{2}$ and the potential energy of the remaining longitudinal mode $\frac{1}{2}$ \cite{Trachenko2016}). On further temperature increase, the system crosses over to the gaslike regime where the phonon phase space continues to reduce, albeit now for longitudinal phonons. In particular, the longitudinal phonons with wavelengths shorter than $\mfp$ disappear because $\mfp$ sets the shortest wavelength in the system. The associated potential energy of the longitudinal phonons reduces, eventually resulting in $c_v=\frac{3}{2}$ as in the ideal gas \cite{Trachenko2016}. Since the phonon energy contributes to liquid $c_V$ \cite{Proctor2020a,proctor1,wallacecv,chen-review,Trachenko2016}, we see that $\ld$ directly affects $c_V$ because it governs the phonon states in the system.

We are grateful to V. V. Brazhkin and J. Proctor for discussions.

\bibliographystyle{apsrev4-1}

\end{document}